\documentclass[pra,twocolumn,showpacs,superscriptaddress,eqsecnum]{revtex4}
\usepackage{amsmath, float}
\usepackage{amsfonts}
\usepackage{amssymb}
\usepackage{graphicx}
\usepackage{bm}
\usepackage{enumerate}
\usepackage{color}
\usepackage{amsthm}
\usepackage{hyperref}

\theoremstyle{plain}
\newtheorem{theorem}{Theorem}
\newtheorem{Proposition}[theorem]{Proposition}

\newtheorem{corollary}[theorem]{Corollary}

\theoremstyle{plain}
\newtheorem{definition}{Definition}

\theoremstyle{plain}
\newtheorem{example}{Example}

\theoremstyle{remark}
\newtheorem{remark}{Remark}

\theoremstyle{protocol}
\newtheorem{protocol} {Protocol}

\begin{document}

\title{Criteria for dynamically stable decoherence-free subspaces and incoherently generated coherences}
\author{Raisa I. Karasik}
\affiliation{Applied Science \& Technology, University of California, Berkeley, California 94720, USA}
\affiliation{Berkeley Quantum Information Center and Department of Chemistry, 
        University of California, Berkeley, California 94720, USA}
\affiliation{Institute for Quantum Information Science,
        University of Calgary, Calgary, Alberta T2N 1N4, Canada}
\author{Karl-Peter Marzlin}
\affiliation{Institute for Quantum Information Science,
        University of Calgary, Calgary, Alberta T2N 1N4, Canada}
\author{Barry C.~Sanders}
\affiliation{Institute for Quantum Information Science,
        University of Calgary, Calgary, Alberta T2N 1N4, Canada}
\affiliation{Centre for Quantum Computer Technology, Macquarie University, Sydney,
        New South Wales 2109, Australia}
\author{K. Birgitta Whaley}
\affiliation{Applied Science \& Technology, University of California, Berkeley, California 94720, USA}
\affiliation{Berkeley Quantum Information Center and Department of Chemistry, 
        University of California, Berkeley, California 94720, USA}

\begin{abstract}
We present a detailed analysis of 
decoherence free subspaces and develop a rigorous
theory that provides necessary and sufficient conditions for dynamically
stable decoherence free subspaces. This allows us to identify a special
class of decoherence free states which rely on incoherent generation
of coherences. We provide examples of physical systems that support such
states. Our approach employs Markovian master equations and applies primarily to finite-dimensional quantum systems.
\end{abstract}

\maketitle
\section{Introduction}
The theory of decoherence-free (or noiseless) subspaces (DFS) provides an
important strategy for the passive preservation of quantum
information~\cite{Dua98, Zan97prl, Lid98,Zan98} and has been subjected to
successful but restricted experimental tests~\cite{Kwi00,
  Kie01,Vio01,Moh03,Lan05, Oller}. DFS theory has been developed in the context of
algebraic symmetries in the interaction Hamiltonian by Zanardi and Rasetti
(ZR)~\cite{Zan97,Zan97prl} and by Lidar, Bacon, and Whaley~\cite{Lid99}, and in the
context of semigroup dynamics via quantum master equations by Lidar,
Chuang, and Whaley (LCW)~\cite{Lid98} and by Shabani and
Lidar~\cite{Shabani05}.  The latter analysis has been extended to the theory of DFS in the non-Markovian regime~\cite{Bac99, Shabani05} and to DFS with 
imperfect initialization~\cite{Shabani05}.
Analogues of DFS have also been proposed in the quantum channel formulation \cite{Zanardi04}.
Potential advantages of DFS include quantum
circuit simplifications by reducing the demands for quantum error correction
and for quantum memory and transport. Given the significance of DFS for
realization of quantum information processing, it is important that the
foundations of the theory are rigorous and unambiguous, especially in
preparation of future experiments to test and exploit the DFS. 

DFS have been defined as collections of states that undergo unitary evolution in the presence of couplings to the environment (i.e., decoherence effects). However, unitary evolution of a quantum state can arise in a number of ways and this fact has resulted in the development of related, but different definitions of DFS in the literature. 
In the context of Markovian master equations, DFS have frequently been defined as a collection of states for which dissipative (decoherence) part of the Markovian master equation is zero.  In this paper, we give a detailed analysis of DFS based on Markovian semigroup master equations defined for finite-dimensional Hilbert spaces. We provide a definition for dynamically stable DFS that identifies a new subclass of DFS states.  This definition allows us to develop a theorem that gives necessary and sufficient conditions for the existence of dynamically stable DFS and that provides a constructive protocol for determining the composition of all DFS. 

Existence of this new class of DFS states is based on the rather counter-intuitive phenomenon that the decoherence term, which is of Lindblad type, can generate coherences between these states and 
other states which are then exactly cancelled by the system Hamiltonian.
This is in striking contradiction to the usual role of the decoherence term, which includes damping of coherences rather than cancellation of these. This new phenomenon is different from techniques to suppress decoherence like dynamical decoupling \cite{Vio98, Zan99, Vio99, Vio99_2} and continuous error correction \cite{Sarovar04, Sarovar05} because it does not rely on the relative time scale for the system evolution and decoherence processes. 
We show that the existence of these incoherently generated coherences results from an invariance of
the Markovian master equation under a specific transformation and 
provide two examples of physical systems that support these special decoherence-free (DF) states. We also indicate that this new phenomenon is rare and does not exist for finite-dimensional systems subject to the common quantum noise processes such as spontaneous emission, thermal excitation and dephasing.

\section{Background}
Our analysis of the DFS is predicated on treating the open system as Markovian and thus using Markovian master equations.

A system~S with state vectors in finite-dimensional Hilbert
space $\mathcal{H}_\text{S}$ is coupled to a reservoir~R
with state vectors in Hilbert space~$\mathcal{H}_\text{R}$.
The system-plus-reservoir (S+R) dynamics are fully described by the Hamiltonian
\begin{equation}
        \hat{H}=\hat{H}_\text{S}\otimes\hat{\openone}_\text{R}+
                \hat{\openone}_\text{S}\otimes\hat{H}_\text{R}+\hat{H}_\text{int}
\end{equation}
with $\hat{H}_\text{S}$, $\hat{H}_\text{R}$ and $\hat{H}_\text{int}$ the
system, reservoir and interaction Hamiltonians, respectively, and $\hat{\openone}$ the
identity operator.  
We are interested in the evolution of the system's reduced state~$\rho$, which is
obtained by tracing over the reservoir's degrees of freedom (subject to appropriate assumptions: no initial correlations between~S and a Markovian~R and weak SR coupling). Dynamics of the reduced density matrix $\rho$ depends on the effective-system Hamiltonian $\hat{H}_\text{eff}=\hat{H}_\text{S}+\hat{\Delta}$, with $\hat{\Delta}$ the Hermitian contribution that stems from the interaction between the quantum system and the reservoir. The evolution of the  reduced density matrix $\rho$ also depends on the decoherence super-operator 
\begin{equation}
  L_\text{D}[\rho] = \frac{1}{2}\sum_{k,l=1}^{N_\text{S}^2-1}a_{kl}
  \left(\left[\hat{F}_k,\rho\hat{F}^\dagger_l\right]
  + \left[\hat{F}_k\rho,\hat{F}_l^\dagger\right] \right),
\label{eq:LD}
\end{equation}
where $A=\left(a_{kl}\right)$ is a
positive semidefinite and time-independent matrix and
$N_\text{S}=\text{dim}(\mathcal{H}_\text{S})$.  
The set of operators
\begin{equation}
        \mathcal{F}=\left\{\hat{F}_k;k=0,\ldots,N_\text{S}^2-1\right\},\;\hat{F}_0\equiv\hat{\openone},
\end{equation}
is a basis for the space of all linear transformations defined on~$\mathcal{H}_\text{S}$.

The evolution of~$\rho$ is given by the master equation
\begin{equation} 
        \dot{\rho} =  -i[\hat{H}_\text{eff},\rho]+L_D[\rho].
\label{eq:master}
\end{equation}
Henceforth, we refer to Eq.~(\ref{eq:master}) as \emph{Markovian open system dynamics}. Eq.~(\ref{eq:master}) is of Lindblad form \cite{Gor76, Lindblad}. This means that Eq.~(\ref{eq:master}) generates a completely-positive trace preserving map and thereby ensures that the solution to the master equation has all the required properties of a physical density matrix at all times.

We note here that the terms
$\hat{\Delta}$ and $L_\text{D}[\rho]$ in Eq.~(\ref{eq:master}) originate from the system-reservoir
coupling and induce a modified unitary evolution and decoherence in system S. In this paper we are interested
in characterizing states that undergo purely unitary time evolution.

\section{Restricted decoherence-free subspaces} \label{sec:restricted}

DFS have been defined as collections of states that evolve in a unitary fashion \cite{Zan97prl, Lid98, Zan98, Zan97, Lid99, Lid01, Shabani05}. 
We shall refer to the complete set of such states as a dynamically stable DFS. From now on we will always assume that the state of the system $\rho(t)$ satisfies Eq.~(\ref{eq:master}).
In this section, we characterize the restricted subset of DFS states $\rho$
fulfilling $L_\text{D}[\rho(t)] = 0$ for all $t$.  Conditions for the more general definition of all states
evolving in a unitary fashion will be considered in Sec.~\ref{secGenDFS}.

The traditional approach to DFS in the context of Markovian master equations
requires that $L_\text{D}[\rho(t)]=0$ \cite{Lid98}.  This constitutes a condition for an instantaneous DFS,
i.e., a DFS at a specific time $t$.  However, 
satisfying the condition $L_\text{D}[\rho(t)]=0$ at one specific time alone, i.e., having an instantaneous DFS, does not
guarantee unitary evolution. The action of the effective system Hamiltonian $\hat{H}_\text{eff}$
modifies the density matrix $\rho(t)$ and thus at a later time $t' > t$, $\rho(t')$ might no longer
describe a pure decoherence-free state.  
The condition
$L_\text{D}[\rho(t)] = 0$ for all $t$ further ensures that state $\rho(t)$ undergoes
pure unitary evolution and thus extends the notion of an instantaneous DFS.  Nevertheless, as we will demonstrate explicitly in Section~\ref{secIGC}, this condition  
is sufficient, but not necessary, for the existence of a DFS.  Therefore we shall refer to the set of states satisfying this sufficiency condition as a restricted DFS. 
\begin{definition}
\label{def:DFS-1}
        Let the time evolution of a state $\rho(t)$ be given by Markovian open system dynamics. Then a state $\rho(t)$ is a pure
        restricted decoherence-free state if $\rho^2(t)=\rho(t)$  
        and $L_\text{D}[\rho(t)]=0$, $\forall\,t$.
\end{definition}
\begin{remark}
\label{re:purestate}
        We require states to be pure in Definition~\ref{def:DFS-1}, i.e., $\rho = |\psi\rangle\langle\psi|$ and
        $\rho^2(t) = \rho(t)$, in order
        to ensure that all states satisfying
        this definition do indeed undergo unitary evolution. For mixed states, it
        is possible to find states that are
        formally decoherence-free (i.e.,  $L_D[\rho(t)]=0$ for all $t$), yet do
        not undergo a proper unitary dynamics. Examples of
        such states include the completely mixed
        state~$\rho(t)=\hat{\openone}_\text{S}/N_\text{S}$ in a system that
        decoheres through dephasing only, as well as many thermal states. The
        reason that these states are not decohering according to Definition~\ref{def:DFS-1} is 
        that the incoherent transition rates 
        between any two states in the mixture are equal, without the dynamics
        of each being unitary. Such states contain minimal
        information and are not useful for the purpose of quantum information
        processing.  
        Hence we focus in this paper on pure states. Our results can be generalized to
        specific mixtures of pure decoherence-free states~\cite{Lid98} but 
        not to a thermal state.
\end{remark}
This definition of a restricted DFS  implies that a pure state  
$\rho(t)$ would remain pure during its
evolution and thus be an ideal tool to process quantum information.
It is therefore important to find criteria for the existence of a restricted DFS in a
given quantum system.  We provide such criteria in Section~\ref{comLCW} below.  Before deriving these criteria, we first discuss the conditions for an instantaneous DFS for which $L_\text{D}[\rho(t)]=0$ at a fixed time $t$, tightening existing conditions that have been derived previously in the Markovian setting~\cite{Lid98}.

\subsection{Instantaneous decoherence-free states}
\label{subsec:existence}
The condition for existence of instantaneous DFS states has been previously
studied by LCW, who formulated the following theorem which we restate here
and then tighten. 
\begin{theorem}
\label{theorem:LCW}
        \textsc{(LCW Theorem~\cite{Lid98})}
        A necessary and sufficient condition for generic decoherence-free dynamics 
        (i.e. $L_\text{D}[\tilde{\rho}]=0$)
        in a subspace ${\{\cal \tilde{H}}=\text{span}[\{|i\rangle\}]$ of the
        register Hilbert space is that all basis states $|i\rangle$ are degenerate
        eigenstates of all the error generators $\mathcal{F}$:
        \begin{equation}
        \label{eq:errorgenerators}
                \hat{F}_k|i\rangle=c_k|i\rangle\;\forall\; k,\;i.
        \end{equation}
        If the error generators $\{\hat{F}_k\}$ can be closed as a semi-simple Lie algebra, then all states
        in ${\cal \tilde{H}}$ are annihilated by all error generators
        $\mathcal{F}$, i.e. $\hat{F}_k|i\rangle=0 |i\rangle\;\forall\;k,\;i$.
\end{theorem}
We first identify and remedy shortcomings corresponding
to the following two examples concerning: i)~zero eigenvalues for the matrix~$\left(a_{kl}\right)$ in Eq.~(\ref{eq:LD}); and ii)~non-zero eigenvalues $c_k$ for the error generators in Eq.~(\ref{eq:errorgenerators}).
These two examples demonstrate the associated difficulties with this description.

\begin{example}
\label{cex:1}
Consider a 3-level system with basis $\{ |0 \rangle , |1\rangle , |2\rangle\}
$ with vanishing Hamiltonian $H_\text{eff}$ and $L_\text{D}[\rho]$ of the form
(\ref{eq:LD}) with $N_\text{S}=3$. 
$A$ is an $8\times 8$ matrix because there are $N_\text{S}^2-1 =8$
error generators in this example. We choose the components of $A$ as
$a_{kl}=1\;$ for $k,l=1,2$ and zero otherwise, so that only the error generators 
$\hat{F}_1=|0\rangle\langle 1|$ and 
$\hat{F}_2=|0\rangle\langle 2|$ are relevant. 
We can rewrite $A$ as
\begin{equation}
        A=B\Lambda B^\dagger,\;
        B=\frac{1}{\sqrt{2}}\begin{pmatrix}\begin{array} {cc}1&1 \\
            1&1\end{array}&O\\O&\sqrt{2}\openone\end{pmatrix}, \; 
        \Lambda=\begin{pmatrix} \begin{array} {cc}2&0 \\ 0&0\end{array}&O\\O&O
        \end{pmatrix}\; ,
\end{equation}
where $O$ denotes sub-matrices with all components
equal to zero and $\openone$ is a $6 \times 6$ identity matrix.
Obviously $A$ possesses zero eigenvalues.
The state $|\psi\rangle= 2^{-1/2} (|1\rangle-|2\rangle)$ is in a DFS 
because
\begin{equation}
        L_\text{D}[|\psi\rangle\langle \psi|]= [H_\text{eff},|\psi\rangle\langle
\psi|]= 0.
\end{equation}
On the other hand we have
$        \hat{F}_{1,2}|\psi\rangle= \pm 2^{-1/2} |0\rangle \neq c_{1,2}|\psi\rangle$,
so that $|\psi\rangle$ is not an eigenstate of the error generators
$\hat{F}_k$, and the LCW theorem would thus not identify $|\psi\rangle$ as a DFS state.
\end{example}

This problem originates from the inclusion of zero eigenvalues of $A$
\footnote{This could be interpreted as constituting ``non-generic" conditions, which were
excluded in LCW theorem~\cite{Lid98}, but it applies in many relevant physical systems and constitutes well-known physical phenomenon such as coherent population trapping~\cite{marzlinCJP07}, which is a basis for spontaneous emission cancellation,
STIRAP, RATOS, and VSCPT; therefore it is best dealt with explicitly in a tightened version of the theorem.}
and can be resolved as follows. Let~$B=(b_{kl})$ be a unitary matrix 
that diagonalizes the matrix $A$ and let $\lambda_k\geq 0$ be 
the positive eigenvalues of
matrix $A$. Then we can define Lindblad operators 
\begin{equation}
        \hat{J}_l=\sum_{k=1}^{N_\text{S}^2-1} b_{lk}\hat{F}_k
\end{equation}
that transform Eq.~(\ref{eq:LD}) to the `diagonalized' decoherence super-operator
\begin{equation}
\label{eq:LD2}
L_\text{D}[\rho]=\frac{1}{2}\sum_{l=1}^M\lambda_l \left([\hat{J}_l,\rho
        \hat{J}_l^\dagger]+[\hat{J}_l\rho, \hat{J}_l^\dagger]\right),
\end{equation}
where we have relabeled the eigenvalues $\{\lambda_l\}$ and operators $\{\hat{J}_l\}$ so that
$\lambda_l>0$ for $l=1,\ldots,M \leq N_\text{S}^2-1$ and $\lambda_l=0$ otherwise. In this form
the zero eigenvalues of $A$ do not contribute to $L_\text{D}[\rho]$. 

In the above example we then have
\begin{equation} 
        L_\text{D}[\rho]=2([\hat{J}_1\, ,\, \rho \, \hat{J}^\dagger_1]
        + [ \hat{J}_1\, \rho \, , \, \hat{J}_1^\dagger ] )
\end{equation} 
with
\begin{equation} 
        J_1=\frac{\hat{F}_1+\hat{F}_2}{\sqrt{2}}=\frac{|0\rangle\langle 1|+|0\rangle\langle 2|}{\sqrt{2}}.
\end{equation} 
Then $\hat{J}_1|\psi\rangle=0|\psi\rangle$. Thus the LCW 
theorem holds if it is modified to require that $|\psi \rangle $ is an eigenstate
of the operators $\{\hat{J}_l\}$ rather than of $\{\hat{F}_k\}$.

\begin{example}
\label{cex:2}
Consider a 2-level system with basis $ \{ |0\rangle , |1\rangle \}$
and $L_\text{D}[\rho]$ of the form (\ref{eq:LD2})
with $M=2$, $\lambda_1=\lambda_2 =1$ and the decoherence operators 
\begin{equation}
        \hat{J}_1 = \left|1\left\rangle\right\langle 1\right|,\,
        \hat{J}_2 = \left|0\left\rangle\right\langle 0\right|+\left|0\rangle\langle 1\right|.
\end{equation}
In order to close the Lie algebra $\mathcal{L}$ generated by operators
$\hat{J}_1$ and $\hat{J}_2$, we need to introduce the operator
$ \hat{J}_3=|0\rangle\langle 1|$.
The sub-algebra spanned by $\hat{J}_3$ forms an Abelian ideal
so that $\mathcal{L}$ is not a semi-simple Lie algebra.
The state $|0\rangle$ is an eigenstate of all decoherence operators,
$\hat{J}_1|0\rangle =0|0\rangle$,
$\hat{J}_2|0\rangle=1 |0\rangle$, 
$\hat{J}_3|0\rangle=0|0\rangle$.
Therefore, according to the 
LCW Theorem, $|0\rangle$ should be decoherence-free. However,
\begin{equation}
        L_\text{D}\left[\left|0\left\rangle\right\langle 0\right|\right]
                =-\frac{\left|0\left\rangle\right\langle 1\right|+\left|1\left\rangle\right\langle 0\right|}{2}\neq 0.
\end{equation}
Hence $|0\rangle$ does not exhibit decoherence-free dynamics, yet it is an
eigenstate of all the error generators. 
 
\end{example}

This example shows that, in addition to excluding zero eigenvalues of $A$, 
another condition for eigenstates of error generators $\hat{J}_l$ corresponding to non-zero
eigenvalues is required in order to tighten the LCW theorem. Here we establish a simple
and complete criterion for identifying whether a given state is an instantaneous DFS 
for a general decoherence process described by a Markovian master
equation. We begin with a definition of the decoherence operator $\hat{\Gamma}$ for a system
having decoherence super-operator $L_\text{D}[\rho]$, Eq.~(\ref{eq:LD2}), of the form
        \begin{equation}
        \label{eq:decoherencematrix}
                \hat{\Gamma} \equiv \sum_{l=1}^M\lambda_l\hat{J}^{\dagger}_l\hat{J}_l.
        \end{equation} 
The eigenvalues of $\hat{\Gamma}$ correspond to the inverse lifetimes of
the corresponding eigenstates.

\begin{Proposition}
\label{prop:corLCW}
        \begin{equation}
        L_\text{D}\left[|\psi\rangle\langle\psi|\right]=0
        \end{equation}
         iff $\hat{J}_l|\psi\rangle=c_l|\psi\rangle$ for all $l=1, \ldots, M$ and $\hat{\Gamma}|\psi\rangle=g|\psi\rangle$,
        where 
        $g=\sum_{l=1}^M \lambda_l |c_l|^2$.  
\end{Proposition}

\begin{proof} First we prove that the conditions in Proposition \ref{prop:corLCW} are sufficient.
Suppose there exists a state $|\psi\rangle$ such that
$\hat{J}_l|\psi\rangle=c_l|\psi\rangle$ and  $\hat{\Gamma}|\psi\rangle=g|\psi\rangle$. 
Then
\begin{align}
        \sum_{l=1}^M \lambda_l \hat{J}_l 
        |\psi\rangle\langle\psi| \hat{J}_l^\dagger = & 
        \left(\sum_{l=1}^M \lambda_l |c_l|^2\right) |\psi\rangle\langle\psi|,
\end{align}
and for $\rho=|\psi\rangle\langle\psi|$
\begin{align}
        L_\text{D}[\rho]=\frac{1}{2}\left(\sum_{l=1}^M 2\lambda_l\hat{J}_l \rho
        \hat{J}^\dagger_l -  
        \rho \hat{\Gamma}-\hat{\Gamma}\rho\right)=0 \;,
\end{align}
because $g=\sum_{l=1}^M \lambda_l |c_l|^2$.

To prove that the conditions in the proposition are necessary, we now assume  $L_\text{D}[\rho]=0$. 
First we show that $\hat{J}_l|\psi\rangle=c_l|\psi\rangle$. 
Using Eq.~(\ref{eq:LD2}) we evaluate the expression
\begin{equation} 
  0 = \langle\psi|\, L_\text{D}\left[\rho \right]\, |\psi\rangle =
   \sum_{l=1}^M \lambda_l \left ( \left|\langle\psi|\hat{J}_l
          |\psi\rangle\right|^2 - \left\|\hat{J}_l |\psi\rangle\right\|^2  \right ).
\label{eq:dfs0}
\end{equation} 
$\hat{J}_l |\psi\rangle$ can generally be written as
$\hat{J}_l|\psi\rangle=c_l|\psi\rangle+|\psi_l^\bot\rangle$
with $|\psi_l^\bot\rangle$ some (non-normalized) state 
that is orthogonal to the state $|\psi\rangle$.
Substituting this into Eq.~(\ref{eq:dfs0}) yields
\begin{equation}
        \sum_{l=1}^M \lambda_l  \langle\psi_l^\bot|\psi_l^\bot\rangle=0.
\end{equation}
Because $\lambda_l>0$ for $l=1,\ldots, M$ we find
$\left\||\psi_\bot\rangle_l\right\|=0$, i.e.,
$\hat{J}_l|\psi\rangle=c_l|\psi\rangle$ $l=1,\ldots, M$.
We also need to show that $\hat{\Gamma}|\psi\rangle=g|\psi\rangle$. 
Consider a subspace
\begin{equation}
        V^\bot=\{|\psi_\bot\rangle\; |\; \langle \psi_\bot|\psi\rangle=0\}.
\end{equation}
Then
\begin{equation}
        0 = \langle\psi_\bot|L_\text{D}[\rho]|\psi\rangle
        =-\frac{1}{2}\langle\psi_\bot|\hat{\Gamma}|\psi\rangle
\;\forall |\psi_\bot\rangle\,\in\,V^\bot.
\end{equation}
This is only possible if $\hat{\Gamma}|\psi\rangle=g|\psi\rangle$.

The value of $g$ can be derived as follows.
Consider the operator
\begin{equation}
        \hat{G}=\sum_{l=1}^M \lambda_l c_l^\ast\hat{J}_l.
\end{equation}
The state $|\psi\rangle$ is an eigenstate of operators
$\hat{G}$ and $\hat{G}^\dagger$ with eigenvalues $g'= \sum_{l=1}^M \lambda_l
|c_l|^2$ and $g$, respectively.  
Then 
\begin{equation}
\label{eq:eig}
       (g')^2=\|\hat{G}^\dagger|\psi\rangle\|^2 =
        \langle\psi|\hat{G}\hat{G}^\dagger|\psi\rangle=g g', 
\end{equation}
assuming that the state $|\psi\rangle$ is normalized. Therefore, $g=\sum_{l=1}^M \lambda_l |c_l|^2$.
\end{proof}
\begin{remark}
        Our Proposition~\ref{prop:corLCW} immediately removes example~\ref{cex:2} 
        because the example does not satisfy the condition
        $\hat{\Gamma}|\psi\rangle=g|\psi\rangle$.
\end{remark}
\begin{corollary}
\label {prop:DFS}
        Suppose that all eigenvalues of operators $\hat{J}_l$  for $l=1,\ldots, M$ 
        are equal to zero; then the state~$|\psi\rangle$ is an instantaneous pure 
        decoherence-free state  iff it is an eigenvector of the 
        decoherence operator~(\ref{eq:decoherencematrix}) with an eigenvalue of zero.
\end{corollary}

We now discuss two examples for which the second condition in Proposition~\ref{prop:corLCW}, 
$\hat{\Gamma} |\psi\rangle = g|\psi\rangle$, is rendered redundant by virtue of the intrinsic physics, i.e., the physical system automatically satisfies this condition.   These examples are drawn from the most common types of noise in quantum   systems, namely excitation, decay and dephasing. 

The first example is noise due to excitation or decay.  During excitation, the system transits to a different state and gains energy. For decay, the system changes its state and looses energy.   For decay processes  the corresponding operators $\hat{J}_l$ can be represented by 
upper triangular matrix with zeros along the diagonal and thus can have only
zeros for its eigenvalues. Excitation processes are correspondingly represented by lower triangular matrices with zeros along the diagonal and therefore also have only zero eigenvalues.  Consequently this kind of noise automatically satisfies Corollary~\ref{prop:DFS} and hence the second condition in Proposition~\ref{prop:corLCW} ( i.e., $\hat{\Gamma}|\psi\rangle=g|\psi\rangle$) is redundant for quantum systems with noise comprised of independent decay and excitation processes. 

The second example where the second condition in Proposition~\ref{prop:corLCW} becomes redundant is pure dephasing.  During dephasing \cite{Sanders}, a quantum system remains in the same state, but acquires a phase.  Thus for quantum noise due to dephasing, there exists a representation such that all $\hat{J}_l$ are diagonal. In this case, $\hat{J}_l$ commutes with $\hat{J}^\dagger_l$, which means that $\hat{J}_l$ is a normal operator \cite{Fried}.   As a result, we know that eigenstates of $\hat{J}_l$ are also eigenstates of $\hat{J}^\dagger_l$ \cite{Fried}. 
Consequently the action of each term contributing to $\hat{\Gamma}$ in Eq.~(\ref{eq:decoherencematrix}) is of the form
$\hat{J}^{\dagger}_l\hat{J}_l|\psi\rangle =  c_l^\ast c_l|\psi\rangle $ with $|\psi\rangle$ a simultaneous eigenstate of 
all $\hat{J}_l$ and $\hat{J}^{\dagger}_l$, so that we automatically have $\hat{\Gamma} |\psi\rangle = g |\psi \rangle$ with $g$ a constant.  Thus in this situation the second condition of Proposition~\ref{prop:corLCW} also becomes redundant. 

At this point it is useful to compare the results of Zanardi~\cite{Zan98} to the LCW theorem and this new tightened version.   In~\cite{Zan98}, Zanardi developed a theorem similar to theorem 3 of LCW~\cite{Lid98}.  Zanardi's theorem applies only to a restricted set  
of systems and does not need tightening. The theorem refers to a quantum system~S that consists of
identical spin-like particles, each of which possesses a finite number of equally spaced energy levels. It is also assumed that this system admits decoherence due to either dephasing, or decay, or excitation, alone and not to a combination of these processes. As we indicated above, the new condition in Proposition
\ref{prop:corLCW} is not required for such independent sources of decoherence.  In general, Theorem~\ref{theorem:LCW} with Proposition~\ref{prop:corLCW} provides tight conditions for the existence of instantaneous DFS in all situations described by a Markovian master equation.

\subsection{Existence criteria for dynamically stable restricted DFS}
\label{comLCW}

In section~\ref{subsec:existence}, we determined when pure states $\rho(t)$
are decoherence-free at a fixed time $t$.  However, as mentioned before, this
is not sufficient for unitary dynamics because the effective system
Hamiltonian $\hat{H}_\text{eff}$ can drive $\rho(t)$ out of the
decoherence-free subspace at a later time $t$. In this section we will
incorporate time evolution in our considerations in order to establish more
precise criteria for the existence of DFS at all times. 
This will characterize restricted DFS that satisfy $L_D[\rho(t)]=0\,\forall\, t$, i.e., dynamically stable DFS. The following theorem  
establishes the composition of ${\cal H}_\text{DFS}$ in terms of all
initial states that undergo decoherence-free evolution. 
\begin{theorem}
\label{th:dfs}
        Let the time evolution be given by Markovian open system dynamics, 
         Eq.~(\ref{eq:master}).
        The space 
        \begin{equation}
                \text{span}\left\{|\psi_1\rangle, |\psi_2\rangle,\ldots,|\psi_K\rangle\right\}
                        =E_\text{Jump}
        \end{equation}
        is a DFS for all time $t$ iff  $E_\text{Jump}$ is a subspace that is invariant 
        under $\hat{H}_\text{eff}$ and that has basis vectors satisfying
        \begin{equation}
                \hat{J}_l|\psi_k\rangle=c_l|\psi_k\rangle,\;
                \hat{\Gamma}|\psi_k\rangle=g|\psi_k\rangle
        \end{equation}
        for all $l=1,\ldots, M$ and for all $k=1,\ldots, K$
        with $g=\sum_{l=1}^M \lambda_l |c_l|^2$ .
\end{theorem}
\begin{proof}
(Note: we have used the notation $E_\text{Jump}$ to emphasize the relation with the operators 
$\hat{J}$ which correspond to the jump operators in a quantum trajectory description (see below).) To prove that conditions of Theorem~\ref{th:dfs} are necessary, we suppose $E_\text{Jump}$ is  a DFS. 
Then any state
\begin{equation}
        |\psi (0) \rangle \in {\cal H}_\text{DFS}
\end{equation}
can be written as
\begin{equation}
        |\psi (0) \rangle = \sum_{k=1}^K \alpha_k |\psi_k \rangle,
\end{equation}
which evolves in time according to
\begin{equation}
        |\psi (t) \rangle = U(t) |\psi (0) \rangle
\end{equation}
with
$       U(t) = \exp (-i t \hat{H}_\text{eff})$
because
$       L_\text{D}\left[|\psi(t) \rangle \langle \psi(t)|\right]=0\; \forall t$.
As 
\begin{equation}
        |\psi (t) \rangle \in {\cal H}_\text{DFS} =E_\text{Jump},
\end{equation}
we can write the evolving state as
\begin{equation}
        |\psi (t) \rangle = \sum_{k=1}^K \alpha_k(t) |\psi_k \rangle .
\end{equation}
As the state
$       |\psi (0) \rangle \in {\cal H}_\text{DFS}$
is arbitrary, this implies that 
\begin{equation}
        \hat{H}_\text{eff}\sum_{k=1}^K\alpha_k |\psi_k\rangle 
                = \sum_{k=1}^K \beta_k\left|\psi_k\right\rangle,
\end{equation}
i.e., the Hamiltonian~$\hat{H}_\text{eff}$ leaves ${\cal H}_\text{DFS}$ invariant.

Since $L_D[|\psi(0)\rangle\langle\psi(0)|]=0$ for all
$|\psi(0)\rangle\in E_\text{Jump}$, by Proposition~\ref{prop:corLCW}
we know that basis states $|\psi_k\rangle$ must be eigenstates of $\hat{J}_l$
and $\hat{\Gamma}$ $\forall\;l=1,\ldots,M$. To complete the proof that the
conditions of Theorem~\ref{th:dfs} are necessary, we have to show that the
eigenvalues $c_l$ and $g$ are equal for 
all basis states $|\psi_k \rangle $. We do this by establishing a 
contradiction. Suppose two arbitrary basis states $|\psi_{k} \rangle $
and $|\psi_{k'} \rangle $ have different eigenvalues:
\begin{equation}
\label{eq:setPr}
        \hat{J}_l|\psi_k\rangle=c_{l,k}|\psi_k\rangle,\;
        \hat{J}_l|\psi_{k'}\rangle=c_{l,k'}|\psi_{k'}\rangle.
\end{equation}
Then the state
\begin{equation}
\label{eq:sup}
        |\psi \rangle =\frac{|\psi_k\rangle + |\psi_{k'}\rangle}{\sqrt{2}}
\end{equation}
is not an eigenstate of $\hat{J}_l$. However, because 
$|\psi \rangle \in {\cal H}_\text{DFS}$ Proposition~\ref{prop:corLCW} implies 
that $|\psi \rangle $ must be an eigenstate. Hence the eigenvalues
must be equal for all basis states.

To prove that the conditions of Theorem~\ref{th:dfs} are sufficient, 
we consider the action of the Liouvillean
\begin{equation} 
 {\cal L}[.] \equiv -i[\hat{H}_\text{eff} \, , \,
. ] + L_\text{D}[.]
\end{equation} 
on a coherence $|\psi_0 \rangle \langle \psi_0^{\prime}|$
between two arbitrary states 
$|\psi_0 \rangle , |\psi_0^{ \prime} \rangle \in
E_\text{Jump}$.
We then have
\begin{eqnarray} 
   L_\text{D}[\, |\psi_0 \rangle \langle \psi_0^{ \prime }|\, ] &=& 
  \sum_l \lambda_l \Big \{ \hat{J}_l 
  |\psi_0 \rangle \langle \psi_0^{  \prime }|  \hat{J}_l^\dagger  
\nonumber \\ & & 
  - \frac{\hat{\Gamma}}{2} |\psi_0\rangle \langle \psi_0^{\prime}|
  -  |\psi_0 \rangle \langle \psi_0^{\prime }|\frac{\hat{\Gamma}}{2}
  \Big \}
\nonumber \\ &=& 0\; ,
\end{eqnarray} 
because the states are eigenstates of $ \hat{J}_l$ and $\hat{\Gamma}$.
Hence the action of ${\cal L}$ can be reduced to
\begin{eqnarray} 
  {\cal L}[\, |\psi_0 \rangle \langle \psi_0^{\prime}|\, ] 
  &=& - i [\hat{H}_\text{eff} \, , \,
  |\psi_0 \rangle \langle \psi_0^{\prime}| \, ]
\nonumber \\ &=&
  - i |\psi_1 \rangle \langle \psi_0^{\prime}|
  + i |\psi_0 \rangle \langle \psi_1^{\prime}| \; ,
\end{eqnarray} 
with $\hat{H}_\text{eff} |\psi_0 \rangle \equiv |\psi_1 \rangle$.
Because $\hat{H}_\text{eff} $ leaves $E_\text{Jump}$ invariant 
we know that $|\psi_1 \rangle , |\psi_1^{\prime} \rangle
\in E_\text{Jump}$. This is a superposition of two coherences
that are of the same type as the original coherence.
Repeated application of ${\cal L}$
and the linearity of $L_\text{D}$ allows us to infer that the action
of $L_\text{D}$ will not contribute to any power ${\cal L}^n [\rho]$.
Consequently, we find for the time evolution of an arbitrary density 
matrix $\rho(0)$ formed of states $\in E_\text{Jump}$,
\begin{equation} 
  e^{t {\cal L}}[\rho(0)] = e^{-i[H_\text{eff}\,, \, .]}\rho(0) = 
  e^{-i t H_\text{eff}} \rho(0) e^{i t H_\text{eff}}\; .
\end{equation} 
Hence, $\rho(t) \in {\cal H}_\text{DFS}$, since it shows unitary evolution under $H_\text{eff}$.
\end{proof}
\begin{remark}
By adopting Definition~\ref{def:DFS-1} for DFS, we were able to
derive that $E_\text{Jump}$ is invariant under action of
$\hat{H}_\text{eff}$ as a consequence of the definition, obviating the 
need to impose this condition as part of the definition itself, as was done in Ref.~\cite{Zan97}.
\end{remark}

Next we provide a Proposition with conditions for $E_\text{Jump}$ that are equivalent to
requiring $E_\text{Jump}$ to be invariant under
$\hat{H}_\text{eff}$, i.e., to one of the conditions in Theorem~\ref{th:dfs} above.  The advantage of these alternative conditions is that it might be
easier to verify them for systems that support DFS with large dimensions. 

\begin{Proposition}
Let $E_\text{Jump}$ be a space spanned by some or all of a collection of
simultaneous degenerate eigenstates of $\hat{J}_l$  with eigenvalues $c_l$ for $l=1,\ldots, M$ that are also eigenstates of 
$\hat{\Gamma}$ 
with eigenvalue $g$. Then $E_\text{Jump}$ is
invariant under $\hat{H}_\text{eff}$ iff  
\begin{equation}
        \left[\hat{H}_\text{eff},
        \hat{J}_l\right]\left|\psi\right\rangle=0\left|\psi\right\rangle
        \text{ for all } l=1, \ldots, M 
\label{eq:con1}
\end{equation}
and
\begin{equation}
        \left[\hat{H}_\text{eff}, \hat{\Gamma}\right]|\psi\rangle=0|\psi\rangle
\label{eq:con2}
\end{equation}
for all $|\psi\rangle\in E_\text{Jump}$.
\label{prop:inv}
\end{Proposition}
\begin{proof}
Let $|\psi\rangle$ be an eigenstate of $\hat{J}_l$ ($l=1, \ldots, M$) and 
$\hat{\Gamma}$, with
respective eigenvalues $c_l$ and $g$. Suppose $\hat{H}_\text{eff}$ leaves
$E_\text{Jump}$ invariant. 
Then $\hat{H}_\text{eff}|\psi\rangle$ is an eigenstate of operators  $\hat{J}_l$ and $\hat{\Gamma}$ with respective eigenvalues $c_l$ and $g$. Consequently, Eqs.~(\ref{eq:con1}) and~(\ref{eq:con2}) hold. 
Suppose now that  Eqs.~(\ref{eq:con1}) and~(\ref{eq:con2}) are true for any $|\psi\rangle\in E_\text{Jump}$. Then
\begin{align}
\hat{J}_l\left(\hat{H}_\text{eff}|\psi\rangle\right)&=c_l\left(\hat{H}_\text{eff}|\psi\rangle\right) \nonumber \\ \hat{\Gamma}\left(\hat{H}_\text{eff}|\psi\rangle\right)&=g\left(\hat{H}_\text{eff}|\psi\rangle\right),
\end{align}
i.e. $\hat{H}_\text{eff}|\psi\rangle$ is in $E_\text{Jump}$ for all $|\psi\rangle\in E_\text{Jump}$.
\end{proof}

\begin{remark}
For the specific examples of decoherence processes describing either dephasing, or decay, or excitation
in a finite-dimensional spin-like systems, it was already shown in Ref.~\cite{Zan98} that the set of states 
\begin{equation}
        \left\{|\psi_1\rangle,|\psi_2\rangle,\ldots, |\psi_K\rangle\right\}
\end{equation}
that fulfills
\begin{equation}
        L_\text{D}\left[|\psi_i\rangle\langle\psi_i|\right]=0
\end{equation}
spans a DFS if it is invariant under $\hat{H}_\text{eff}$.
\end{remark}.

\section{General conditions for dynamically stable Decoherence-free subspaces}\label{secGenDFS}
In the previous section we have studied pure, restricted DFS that fulfill
the sufficiency criterion of Definition~\ref{def:DFS-1}, namely $L_\text{D}[\rho(t)]=0$ for all $t$. However,  for unitary dynamics, it is only necessary that the purity $\text{Tr}[\rho^2(t)]$ of a state is preserved during the evolution 
(given an initially pure state, i.e., with $\text{Tr}[\rho^2(0)]=1$). This motivates us to provide a more general definition for the DFS as a collection of states that evolves in a unitary fashion. This is a larger set than the restricted DFS studied in Section~\ref{sec:restricted}.

Let $D(\mathcal{H})$ be the set of all density matrices that describe pure states and that are  defined for a quantum system associated with the Hilbert space $\mathcal{H}$.
\begin{definition}
Let the time evolution of an open quantum system with Hilbert space
${\cal H}_\text{S}$ be given by Markovian open system dynamics, 
Eq.~(\ref{eq:master}).
Then a decoherence free subspace $\mathcal{H}_\text{DFS}$ is a subspace of
$\mathcal{H}_\text{S}$ such that all pure states $\rho(t)\in
D(\mathcal{H}_\text{DFS})$  fulfill 
\begin{equation}
 \partial_t\text{Tr}[\rho^2(t)]=0 \, \forall\,t\geq 0, \text{ with }\text{Tr}[\rho^2(0)]=1.
\end{equation}
\label{def:gDFS}
\end{definition}
We show in Appendix \ref{app:pure} that all states statisfying Definition \ref{def:gDFS} undergo unitary evolution and therefore constitute a true DFS.

\begin{remark}
In Definition \ref{def:gDFS}, it is absolutely crucial that the purity of the state $\rho(t)$ is preserved for all times $t$, since non-unitary evolution is possible at intermediate times even when the initial state and final states are pure. For example, consider a 2-level atom subject to decay in the upper level by spontaneous emission.  For this system, an initially pure excited state will eventually evolve into a pure ground state. Thus, initial and final states are both pure.  However, for any intermediate time between 
$t=0$ and $t \rightarrow \infty$, the state of the system will be an incoherent mixture of ground and excited states. Therefore, despite starting out as a pure excited state, this system will not remain pure at all times and will therefore not be classified as DFS by Definition \ref{def:gDFS}.  
\end{remark}

Using Definition~\ref{def:gDFS} and Eq.~(\ref{eq:master}), it immediately follows that \cite{lidarSchneider2005, LSA}
\begin{equation}
\partial_t\text{Tr}[\rho^2(t)]=2\langle L_\text{D}[\rho(t)] \rangle.
\end{equation}
Therefore, DFS exist when $\langle L_\text{D}[\rho] \rangle=0$,
which is less restrictive than $ L_\text{D}[\rho]=0$. We will analyze
the difference between these two conditions in Sec.~\ref{secIGC}.
In the remainder of this section
we  present a theorem that gives necessary and sufficient conditions for the existence of DFS
according to Definition~\ref{def:gDFS} 
and also provide a constructive protocol for determining both when DFS exist and the exact specification of all states within the DFS.
\begin{theorem}
\label{th:dfs2}
        Let the time evolution of an open quantum system described in
        a
        finite-dimensional Hilbert space be governed by 
        Eq.~(\ref{eq:master}) with time-independent $\hat{H}_\text{eff}$. 
        The space 
        \begin{equation}
                \text{span}\left\{|\psi_1\rangle,
                  |\psi_2\rangle,\ldots,|\psi_K\rangle\right\} 
                        =E_\text{Jump}
        \end{equation}
        is a DFS satisfying Definition~\ref{def:gDFS} iff the basis vectors fulfill
        \begin{equation}
                \hat{J}_l|\psi_k\rangle=c_l|\psi_k\rangle
        \label{eq:cond1}
        \end{equation}
        for all $l=1,\ldots, M$ and for all $k=1,\ldots, K$ and
        $E_\text{Jump}$ is invariant  
        under 
        \begin{equation}
        \hat{H}_\text{ev}=\hat{H}_\text{eff}+\frac{i}{2}\sum_{l=1}^M
        \lambda_l\left(c^\ast_l\hat{J}_l-c_l\hat{J}_l^\dagger \right).
        \label{eq:cond2}
        \end{equation}
        This guarantees identification of all DFS as well as explicit construction of all DF states. 
\end{theorem}

\begin{proof}
Before proceeding with the proof, we point out that $E_\text{Jump}$ contains only those eigenvectors of $\hat{J}_l$ that span an invariant subspace with respect to $\hat{H}_\text{ev}$, i.e., the dimension of  $E_\text{Jump}$ may be  equal to or less than the number of simultaneous linearly independent eigenvectors of the jump operators $\hat{J}_l$ ($l=1, \ldots, M$) and there is no requirement in the theorem that these be the same.

We begin the proof by observing that for any $M$-dimensional vector $\vec{b}$ with complex
coefficients $b_1, \cdots, b_M$ \cite{Breuer}
\begin{equation}
L_D[\rho]=\widetilde{L}_D^{(\vec{b})}[\rho]-i\left[\frac{i}{2}\sum_{l=1}^M
\lambda_l\left(b^\ast_l\hat{J}_l-b_l\hat{J}_l^\dagger \right) , \rho\right], 
\label{eq:transformation}
\end{equation}
where
\begin{equation}
\widetilde{L}_D^{(\vec{b})}[\rho]=\frac{1}{2}\sum_{l=1}^M\lambda_l
\left(\left[\tilde{J}_l(b_l),\rho 
        \tilde{J}_l^\dagger(b_l)\right]+\left[\tilde{J}_l(b_l)\rho, \tilde{J}_l^\dagger(b_l)\right]\right)
        \label{eq:transformation_2}
\end{equation}
with $\tilde{J}_l(b_l)=\hat{J}_l-b_l\hat{I}$.

Now we prove that the conditions of Theorem~\ref{th:dfs2} are sufficient. 
Any state
\begin{equation}
        |\psi (t) \rangle \in E_\text{Jump}
\end{equation}
can be written as
\begin{equation}
        |\psi (t) \rangle = \sum_{k=1}^K \alpha_k(t) |\psi_k \rangle \text{
          and }\rho(t)=|\psi(t)\rangle\langle\psi(t)|. 
\end{equation}
Then for $\vec{c}=(c_1, \cdots, c_M)$ 
\begin{eqnarray} 
  L_D[\rho(t)] &=& \widetilde{L}_D^{(\vec{c})}[\rho]-i[\frac{i}{2}\sum_{l=1}^M
  \lambda_l\left(c^\ast_l\hat{J}_l-c_l\hat{J}_l^\dagger \right) ,
  \rho]
\nonumber \\ &=&
  -i[\frac{i}{2}\sum_{l=1}^M
  \lambda_l\left(c^\ast_l\hat{J}_l-c_l\hat{J}_l^\dagger \right) , \rho] 
\label{eq:dec}
\end{eqnarray} 
because
$\tilde{J}_l(c_l)|\psi(t)\rangle=(\hat{J}_l-c_l\hat{I})|\psi(t)\rangle=0|\psi(t)\rangle$. 
Hence the time evolution for $\rho(t)$ is given by
\begin{equation}
\dot{\rho}(t)=-i\left[\hat{H}_\text{ev}, \rho(t)\right]
\end{equation}
so that
\begin{align}
\partial_t\text{Tr}[\rho^2(t)]&=2\text{Tr}[\dot{\rho}(t)\rho(t)]\nonumber \\
&=2\text{Tr}\left[-i[\hat{H}_\text{ev}, \rho(t)]\rho(t) \right]=0.
\end{align}
Thus  $E_\text{Jump}$ is  a DFS. 

To prove that the conditions are necessary we suppose $E_\text{Jump}$ is a DFS. 
For fixed $t_0$,
\begin{equation}
        |\psi (t_0) \rangle = \sum_{k=1}^K \alpha_k(t_0) |\psi_k
         \rangle,\,\,\,\rho(t_0)=|\psi(t_0)\rangle\langle\psi(t_0)|, 
\end{equation}
which implies
\begin{eqnarray} 
\label{eq:exVforLD}
  0 &=& \partial_t\text{Tr}\left[\rho^2(t_0)\right]
\nonumber \\ &=&
   2\text{Tr}\left[\rho(t_0)L_D[\rho(t_0)]\right]
\nonumber \\ &=&
   \langle\psi(t_0)|L_D[\rho(t_0)]|\psi(t_0)\rangle \; .
\end{eqnarray} 
Suppressing the dependence on $t_0$ we
see that Eq.~(\ref{eq:exVforLD}) has the same form as Eq.~(\ref{eq:dfs0}). Repeating
the same argument we are led to the condition $\hat{J}_l|\psi\rangle=c_l|\psi\rangle$. 
Furthermore, the line of reasoning presented in Eqs.~(\ref{eq:setPr}) and
(\ref{eq:sup}) proves that the eigenvalues $c_l$ are equal for 
all basis states $|\psi_k \rangle $. Thus we can conclude that
$\hat{J}_l|\psi(t_0)\rangle=c_l(t_0)|\psi(t_0)\rangle$, for
any $t_0$.  

At this point, it appears that the eigenvalue $c_l(t)$ may depend on
time $t$, but we argue now that $c_l(t)$ must be time-independent for
finite-dimensional systems, i.e. $c_l(t)=c_l$. Eq.~(\ref{eq:master}) is a
system of 
linear, first-order
differential equations with constant coefficients; the components of
$\rho(t)$ and  $|\psi(t)\rangle$ are therefore continuous functions in time.
$\hat{J}_l|\psi(t)\rangle=c_l(t)|\psi(t)\rangle$ then implies that $c_l(t)$ 
must be continuous, too. On the other hand, 
the time independent operator $\hat{J}_l$ acts on
a finite-dimensional Hilbert space so that its spectrum is discrete. Thus, $c_l(t)$
must be constant.   

According to Eq.~(\ref{eq:dec}) for $\rho(t)=|\psi(t)\rangle\langle
\psi(t)|$ with $\hat{J}_l|\psi(t)\rangle=c_l|\psi(t)\rangle$, we then have
\begin{equation}
 \dot{\rho}(t) =  -i[\hat{H}_\text{eff},\rho(t)]+L_\text{D}[\rho(t)]=-i[\hat{H}_\text{ev},\rho(t)]
\end{equation}
and
\begin{equation}
\rho(t)=\exp(-i\hat{H}_\text{ev}t)\rho(0)\exp(i\hat{H}_\text{ev}t) 
\end{equation} 
with $|\psi(t)\rangle=\exp(-i\hat{H}_\text{ev}t)|\psi(0)\rangle $ because
$\hat{H}_\text{ev}$ is time-independent. 
Thus $|\psi(t)\rangle\in E_\text{Jump}$ if $E_\text{Jump}$
is invariant under $\hat{H}_\text{ev}$. 
The Hilbert space $\mathcal{H}_S$ can in principle contain multiple DFS. Each DFS is characterized by an $M$-tuple $(c_1, \ldots, c_M)$ of eigenvalues of all the jump operators $\hat{J}_1, \ldots, \hat{J}_M$ for states in $E_\text{Jump}$.

Now we demonstrate that Theorem \ref{th:dfs2} is constructive, i.e., we explain how to use Theorem \ref{th:dfs2} to determine all basis states in all DFS. Condition 1 of Theorem \ref{th:dfs2}, i.e.,  Eq.~(\ref{eq:cond1}), states that DFS are subsets of all common degenerate eigenstates of the jump operators $\hat{J}_1, \ldots, \hat{J}_M$. We label the space of all common degenerate eigenstates of the jump operators as $E_J$, with dimension $N$.  
The space $E_J$ is determined by a collection of corresponding eigenvalues of the jump operators, $c_1, \ldots, c_M$. A different collection of eigenvalues might give rise to a different nonempty space $E_J$ and constitute potentially a different DFS.   In order to satisfy condition 2 of Theorem \ref{th:dfs2}, i.e., invariance under Eq.~(\ref{eq:cond2}), we need to restrict the space $E_J$ to a subspace $E_{Jump}$  that is invariant under $\hat{H}_\text{ev}$.  We accomplish this task using Proposition~\ref{prop:inv}. Proposition \ref{prop:inv} says that a space that is spanned by some or all of the collection of degenerate simultaneous eigenstates of a collection of operators ($\hat{J}_1, \ldots, \hat{J}_M$ in this case), is invariant under the action of a Hamiltonian $\hat{H}$ iff all states in the space are eigenstates of the corresponding commutator operators $B_l=[\hat{H}, \hat{J}_l]$ with eigenvalues 0, for all $l=1, \ldots, M$. Thus, in order to construct DFS from $E_J$, we i) consider all basis elements $|\psi_1\rangle, \ldots, |\psi_N\rangle $ for $E_J$, ii) form linear combinations of these as $\alpha_1|\psi_1\rangle+\cdots+\alpha_N|\psi_N\rangle$ with $\alpha_1, \ldots, \alpha_N$ free parameters, and iii) then determine all possible sets $\{\alpha_i, i=1, \ldots N \}$ such that the resulting states are eigenstates of $\hat{B}_l=[\hat{H}, \hat{J}_l]$ with eigenvalue 0 for all 
$i=1, \ldots, M$. The resulting states form a complete basis for the DFS  specified by  eigenvalues $c_1, \ldots, c_M$ of the jump operators. Different choices of eigenvalues for the jump operators can give rise to different DFS and are constructed similarly.  
\end{proof}

\begin{remark}
Theorem \ref{th:dfs2} applies only to finite-dimensional
spaces. To demonstrate this we consider a damped harmonic oscillator described by the master equation
\begin{equation}\dot{\rho}(t)=-i[\omega_0\hat{a}^\dagger\hat{a},
  \rho(t)]+\frac{\gamma}{2}[2\hat{a}\rho(t)\hat{a}^\dagger-\hat{a}^\dagger\hat{a}\rho(t)
  -\rho(t)\hat{a}^\dagger\hat{a}] ,
\end{equation}
with $\hat{a}$ the annihilation operator. 
A coherent state $\rho(t)=|\alpha(t)\rangle\langle\alpha(t)|$, with
$\alpha(t)=\alpha \, \text{e}^{-(i\omega_0-\gamma/2)t}$ and $\alpha$ a
complex constant, is a solution to this master equation. $|\alpha(t)\rangle$ is an eigenstate
of $\hat{a}$ with unit purity, but it is not invariant under the evolution Hamiltonian
$\hat{H}_\text{ev}(t)=\omega_0\hat{a}^\dagger\hat{a}+\frac{i\gamma}{2}(\alpha^\ast(t)\hat{a}-\alpha(t)\hat{a}^\dagger)$.
Theorem \ref{th:dfs2} does not apply to this example because a coherent state
$|\alpha(t)\rangle$ is an eigenstate of $\hat{a}$ with a time-dependent
eigenvalue $\alpha(t)$. Thus, the analysis following Eq.~(\ref{eq:exVforLD}) 
in the proof of Theorem \ref{th:dfs2} does not apply. 
\end{remark}

We now provide a constructive protocol for DFS, i.e., we present a procedure that explicitly determines the existence and composition of all DFS.
\begin{protocol}

\begin{enumerate}
	\item We consider all Lindblad operators, $\hat{J}_l$ $l=1, \ldots, M$, from Eq.~(\ref{eq:LD2}) and calculate common eigenstates for these operators, i.e.,  eigenstates $|\psi_i\rangle$ satisfying $\hat{J}_l|\psi_i\rangle=c_l|\psi_i\rangle, \text{ for } l=1, \ldots, M$.  Note for each $l$, all $|\psi_i\rangle$ are eigenstates of $\hat{J}_l$ with the same eigenvalue. However, eigenvalues for different index $l$ need not be the same.
	The number of such common eigenstates is $N \geq 0$. For $N =0$, there is no DFS.
	\item When $N > 0$ we construct a space $E_J=\text{span}\{|\psi_i\rangle, 
	\,i=1, \ldots, N\}$.  This space is characterized by a set of eigenvalues $c_1, \ldots, c_M$ of the jump operators.  According to Theorem \ref{th:dfs2}, any DFS must  be a subspace of $E_J$. 
	Note that according to Theorem \ref{th:dfs2}, $\text{Dim}(\mathcal{H}_\text{DFS})\leq N$.
	\item Now we apply Proposition \ref{prop:inv} to the conditions determined by Theorem \ref{th:dfs2} to select out the states in $E_J$ that belong to a DFS (note that there is no restriction to a single DFS). 
		The conditions of Theorem \ref{th:dfs2} require that the DFS be invariant under $H_\text{ev}$.  According to Proposition \ref{prop:inv}, a state $|\psi_i\rangle$ in $E_J$  that is invariant under 
		some Hamiltonian $\hat{H}$ is a zero eigenstate of the operator $[\hat{H}, \hat{J}_l]$. 
		Thus, the DFS is a collection of simultaneous eigenstates of $\hat{J}_l$ that are also eigenstates of all $B_l=[\hat{H}_\text{ev}, \hat{J}_l]$, with eigenvalue 0, i.e.,  
		$\mathcal{H}_\text{DFS}=\{|\psi_i\rangle \,: |\psi_i\rangle\in E_J \text{ and } \hat{B}_l |\psi_i\rangle=0 |\psi_i\rangle \} $.
	We can therefore simply construct the operators $\hat{B}_l=[\hat{H}_\text{ev}, \hat{J}_l]$ and use these to select out the states $|\psi_i\rangle$ in $E_\text{Jump}$ that also fulfill $\hat{B}_l |\psi_i\rangle=0|\psi_i\rangle$.	Doing this for all $l$ gives the basis for the DFS that is specified by eigenvalues $c_1, \ldots, c_M$, of total dimension 
	$\text{Dim}(E_\text{Jump})\leq N$.  
	\end{enumerate}
\end{protocol}	
This protocol must be repeated for all possible combinations of common eigenvalues of $\hat{J}_1,\hat{J}_2, \ldots, \hat{J}_M$. This will generate all existing DFS.  
 These results provide simple constructive criteria for the existence of DFS and for exact specification of {\em{all}} DFS states for any given Markovian master equation
\footnote{  For the Markovian setting Ref. \cite{Shabani05} presented an alternative definition of a DFS and, for a given partition of the system Hilbert space and a specific representation, determined conditions on the jump operators and effective system Hamiltonian for a given subspace to be DF  (Theorem 3 in Ref. \cite{Shabani05}); this theorem can be used to verify the DFS determined by Theorem \ref{th:dfs2} but unlike Theorem \ref{th:dfs2}, it does not provide a constructive method for determining the states in the DFS.}.

\section{Incoherently generated coherences}
\label{secIGC}
In this section we analyze the difference between the two definitions of
decoherence free subspaces introduced in this paper, i.e., Definitions~\ref{def:DFS-1} and~\ref{def:gDFS}.
The following example demonstrates that there are states that
satisfy Definition \ref{def:gDFS} but do not satisfy the criteria of Definition
\ref{def:DFS-1}.
\begin{example}
Consider a two-level system with  basis states $|1\rangle$ and $|0\rangle$ in presence of a single decoherence operator
$\hat{J} = \sigma_+ + \sigma_z$, $\lambda=2$, and Hamiltonian $H= \sigma_y$. 
Then the state $|\psi \rangle  =|1\rangle$ has non-zero action of $L_\text{D}$ on it, with
 $L_\text{D}[\, |\psi \rangle \langle \psi|\,] 
= i[\sigma_y , |\psi \rangle \langle \psi|]= -\sigma_x \neq 0$ , and yet it
corresponds to a stationary solution of the master equation that
fulfills $\langle L_\text{D}[\rho] \rangle =0$.
\label{ex:igc}
\end{example}

In absence of the Hamiltonian $H$, the super-operator 
$L_D[\, |\psi \rangle \langle \psi|\,]$ would therefore evolve the
system into a state that differs from $|\psi \rangle $ and that would in general therefore be subsequently affected by decoherence. However in this example, $H$ exactly cancels the effect of  the decoherence super-operator at $t=0$ and hence for all times, ensuring that $|\psi \rangle$ remains
stationary and decoherence-free.  

The reason why $|\psi \rangle $ is stable in this example is that
$L_\text{D}[\rho]$ initially acts coherently on $|\psi \rangle $
and the coherences generated by this are canceled by the action of the unitary term
$-i[H,\, |\psi \rangle \langle \psi|\,]$.   It is easy to see that the initial coherence is generated between the state $|\psi\rangle = |1\rangle$ and the orthogonal state $|0\rangle$.  However this cancellation is specific to these two states and does not hold for an arbitrary state of the two-level system, for which additional terms of the form of Eq.~(\ref{eq:transformation_2}) will be present in the action of $L_\text{D}[\rho]$.  Consequently, if the Hamiltonian term were absent,
$L_\text{D}[\rho]$ would generate an incoherent evolution of the
quantum state $|\psi \rangle $ because it only acts coherently on this particular state and its orthogonal complement, so that
once the initial state has been infinitesimally coherently transformed to some superposition of these states, $L_\text{D}[\rho]$ will gain some contribution of incoherent terms of the form of Eq.~(\ref{eq:transformation_2}).

The existence of states that satisfy Definition \ref{def:gDFS} but not Definition
\ref{def:DFS-1} is surprising because it disagrees with the following,
physically intuitive argument.
Inserting the decoherence operator $\hat{\Gamma}$ 
into Eq.~(\ref{eq:LD2}) we can rewrite the master equation Eq.~(\ref{eq:master}) as
\begin{align}
        \dot{\rho}=&
        -i \hat{H}_\text{nh}\rho + i \rho  \hat{H}_\text{nh}^\dagger 
        +\sum_{l=1}^M \lambda_l \hat{J}_l \rho\hat{J}_l^\dagger
\label{Eq:generalME}
\end{align}
with the non-Hermitian Hamiltonian 
$ \hat{H}_\text{nh} \equiv \hat{H}_\text{eff} -i \hat{\Gamma}/2$.
In the context of quantum trajectory methods
\cite{carmichael93}, $ \hat{H}_\text{nh}$ generates a continuous evolution
in time, while  $\sum_{l=1}^M \lambda_l \hat{J}_l \rho\hat{J}_l^\dagger$ is
interpreted as generating sudden quantum jumps at random times.
The dynamics in absence of quantum jumps is given by
\begin{equation}
  \rho(t)= e^{-i t \hat{H}_\text{nh}}\rho(0)  
 e^{i t \hat{H}_\text{nh}^\dagger }\; .  
\end{equation}
Since $\hat{\Gamma}$ and $\hat{H}_\text{eff}$ are both Hermitian operators, 
they correspond to the anti-Hermitian and Hermitian parts of
$ \hat{H}_\text{nh}$, respectively.
Hence, $\hat{H}_\text{eff}$ cannot eliminate the effect of the
decoherence operator $\hat{\Gamma}$. Naively, one also would expect
that the quantum jump term
$\sum_{l=1}^M \lambda_l \hat{J}_l \rho\hat{J}_l^\dagger$ cannot
cancel the effect of $\hat{\Gamma}$.  This argument leads to the conclusion that setting
$L_\text{D}[\rho(t)]=0$ $\forall\,\, t$ is the only way to achieve unitary
evolution. Example \ref{ex:igc} demonstrates that this argument
is not correct in general.

The existence of these states 
can be quite generally explained with reference to the
properties of Markovian master equations.
Eq.~(\ref{eq:transformation}) shows that the Markovian master equation,
Eq.~(\ref{eq:master}), is invariant under the following transformation
\cite{Breuer}
\begin{align}
\hat{J}_l&\rightarrow\hat{J}_l-b_l \hat{I} \text{ for }l=1, \ldots, M
\\
\hat{H}_\text{eff}&\rightarrow\hat{H}_\text{eff}+\frac{i}{2}\sum_{l=1}^M
\lambda_l \left( b_l^\ast\hat{J}_l-b_l \hat{J}_l^\dagger\right).
\end{align}
Here $b_l$ ranges over all complex numbers.

This transformation implies that $L_D[\rho]$ can always rewritten as a sum
of two terms (see Eq.~(\ref{eq:transformation})), namely, i) a decoherence
super-operator with respect to the transformed jump operators $\hat{J}_l-b_l \hat{I}$,
 and ii)
a commutator between $\rho$ and the Hermitian operator
$\hat{H}_D\equiv\frac{i}{2}\sum_{l=1}^M \lambda_l \left(
b_l^\ast\hat{J}_l-b_l \hat{J}_l^\dagger\right)$. Thus, the 
general decoherence super-operator $L_D[\rho]$ always contains
a commutator with
a Hermitian operator and, therefore, can in principle generate
unitary evolution 
under suitable circumstances.
Consequently, we cannot simply interpret $i[\hat{H}_\text{eff},\rho]$ and
$L_\text{D}[\rho]$ in the Markovian master equation,
Eq.~(\ref{eq:master}), as terms that {\em always} generate unitary and decohering
dynamics, respectively. 

Under special circumstances, in particular, when the state $\rho$ is generated
by a common eigenstate of all jump operators $\hat{J}_l$, Eq.~(\ref{eq:transformation_2}) vanishes and
the decoherence
super-operator consists then of just the commutator with Hermitian $\hat{H}_D$: 
\begin{equation}
 L_D[\rho]=-i[\hat{H}_D ,\rho].
\end{equation}
In this situation, a state $\rho$ decoheres because the Hermitian Hamiltonian $\hat{H}_D$
generated by $L_D[\rho]$ drives state $\rho$ out of the subspace where its purity is
preserved, and not because $L_D[\rho]$ causes a genuine decay in a quantum
system. It is then possible that the effective system Hamiltonian $\hat{H}_\text{eff}$ might compensate for such unitary leakage, resulting in preservation of the subspace. This is the situation 
described in Example \ref{ex:igc} above.    
For a given system, it may therefore be possible to construct or to modify the effective Hamiltonian specifically to ensure 
negation of the effects of $\hat{H}_D$ and to thereby stabilize states $\rho$ in the subspace corresponding to one or more common eigenstates of the jump operators $\hat{J}_l$.

Example  \ref{ex:igc}  thus illustrates an intriguing general phenomenon, i.e., that 
even though decoherence in an
open system S is caused by the interaction between system S and its reservoir,
decoherence can be completely eliminated by a particular system Hamiltonian
under special circumstances. We would like to emphasize here that this effect
is very different from
well-studied 
decoherence suppression phenomena such as dynamical decoupling \cite{Vio98,Zan99, Vio99, Vio99_2}. In dynamical
decoupling, decoherence effects are suppressed because the time scale for
the system Hamiltonian is much faster than the timescale for the decay
processes. The system Hamiltonian changes the state of the system so rapidly
that decoherence processes are adiabatically eliminated.  In contrast, in Example~\ref{ex:igc} the decoherence effects simply do not take place, as in any DFS. 

To understand 
more generally
under which conditions the incoherent part $L_\text{D}[\rho]$
of the master equation can support a DFS through the generation
of coherences, we introduce now the following special class of  DFS states. 
\begin{definition}
\label{def:IGC}
States with incoherent generation of coherences (IGC) are pure states $\rho(t)$
that evolve unitarily, $\partial_t\text{Tr}[\rho^2(t)]=0 \, \forall\,t\geq 0$,
but that have $L_D[\rho(t)]\neq0$ at some time(s) t. \
\end{definition}

Theorem \ref{th:dfs2} allows us to investigate the difference between
restricted DFS states in the sense of Definition \ref{def:DFS-1} and subspaces composed of IGC states, which we shall refer to as IGC subspaces.
Assume state $\rho = |\psi \rangle \langle \psi|$ fulfills 
$\hat{J}_l |\psi \rangle = c_l |\psi \rangle $ so that
$\langle L_\text{D}[\rho] \rangle =\partial_t \text{Tr}[\rho^2(t)]= 0$. 
Then
\begin{equation} 
   L_\text{D}[\rho] = \sum_l \left (
    2 |c_l|^2 \rho - c_l J_l^\dagger \rho - c_l^* \rho J_l 
   \right )\; .
\end{equation} 
The following result can be inferred from this equation.
\begin{corollary}
\label {prop:DFS2}  
  An IGC state $|\psi \rangle $ can only exist if, for at least one $l$, 
  $\hat{J}_l|\psi \rangle = c_l |\psi \rangle $ with
  $c_l \neq 0$, and $|\psi \rangle $ is not an eigenstate of 
  $\sum_l c_l\hat{J}_l^\dagger $.
\end{corollary}

\begin{remark}
This result implies that IGC states and IGC subspaces are rare in nature. In particular, IGC states do not
exist
when the operators $\hat{J}_l$ are normal ($[\hat{J}_l,\hat{J}^\dagger_l]=0$~\cite{Fried}) since in this case any $|\psi \rangle \in {\cal
  H}_\text{DFS}$ would necessarily also be an eigenstate of $\hat{J}_l^\dagger
$. This implies that IGC states do not exist in systems subject to decoherence
due to dephasing (see also the discussion below Corollary 3). In 
decoherence processes that involve population exchange, such 
as spontaneous emission or thermal excitation, a DFS state
typically corresponds to an eigenstate of $\hat{J}_l$ with $c_l=0$, as discussed earlier, and so Corollary~\ref{prop:DFS2} precludes any IGC in this situation. Thus IGC states do not exist in systems with decoherence due to independent decay, excitation or dephasing processes.
Hence, while the intuitive physical idea that 
$ L_\text{D}[\rho]  =0$ should be fulfilled for DFS states is
mathematically wrong, it nevertheless is valid for many
physical applications.
\end{remark}

\subsection{Examples of IGC states}

While IGC states do not exist for the most common physical
decoherence models, they nevertheless can occur under special
circumstances as the following examples illustrate.
\begin{example}Driven two-level atom in squeezed vacuum reservoir\end{example}
We consider a two-level atom that interacts
with a radiation field. We also assume that the radiation field is prepared in a squeezed vacuum state.
Then this system  is described by the master equation \cite{Breuer} 
\begin{align}
  \dot{\rho}=&-i[\hat{H}_S, \rho] 
\label{eq:sq} \\
  &-\frac{\gamma_0}{2} c^2 \left(\sigma_+\sigma_-\rho+\rho\sigma_+\sigma_- - 2\sigma_-\rho\sigma_+\right)
\nonumber \\
  &-\frac{\gamma_0}{2} s^2 \left(\sigma_-\sigma_+\rho+\rho\sigma_-\sigma_+-2\sigma_+\rho\sigma_-\right)
\nonumber \\
  &+\gamma_0\big( s c \sigma_- \rho \sigma_- 
         +s c\sigma_+\rho\sigma_+ \big )\; ,
\nonumber 
\end{align}
with $s \equiv\sinh(r)$, $c \equiv \cosh(r)$ and  $r$ the (real) squeezing parameter. Eq.~(\ref{eq:sq}) assumes Lindblad form when we introduce the operator 
\begin{equation}
\hat{J}= c \sigma_- + s \sigma_+.
\end{equation}
For $r\neq 0$ the operator $\hat{J}$ has two non-zero eigenvalues
$\pm\sqrt{s c}$ with eigenstates $|\psi_\pm \rangle =  \pm \sqrt{s} |0 \rangle
+ \sqrt{c} |1 \rangle $.
Each corresponds to a stationary IGC state if the Hamiltonian is given by
$\hat{H}_S = \pm \frac{\gamma_0}{2}\sqrt{s c}(s-c) \sigma_y $, i.e., the DFS (IGC subspaces) are
at most one-dimensional.

However, one state is not sufficient to encode a qubit. Next we will explore
how one can generate an IGC subspace large enough to be used for quantum
information processing. 

\begin{example} N two-level atoms in the Dicke limit in squeezed vacuum reservoir \end{example}

We consider $N$ two-level atoms in a squeezed vacuum reservoir in the Dicke
limit~\cite{Dic54}. 
We model the dynamics of this system by the master equation 
\begin{equation}
\dot{\rho}=-i[\hat{H}_D, \rho]
+\frac{\gamma}{2}[2\hat{J}\rho\hat{J}^\dagger-\hat{J}^\dagger\hat{J}\rho-\rho\hat{J}^\dagger\hat{J}]
\end{equation}
with 
\begin{equation}
\hat{J}=\sum_{n=1}^N\hat{J}_n  \text{  and   } \hat{J}_n=c\hat{\sigma}_{n-}+s\sigma_{n+},
\end{equation}
where the index $n$ refers to the $n$th atom.

To find IGC states, we need to construct all eigenstates of the operator
$\hat{J}$. They are given by all possible tensor products of 
$|\psi_\pm \rangle$. For a given eigenstate, let $n_+$ be
number of $|\psi_+ \rangle$ components and  $n_-=N-n_+$ the number of
$|\psi_- \rangle$ components. Then the corresponding eigenvalue is
$(n_+-n_-)\sqrt{s c}$. For  $n_+=n_-$ the eigenvalue is zero, indicating
that the state is not of IGC type. All
other eigenstates of $\hat{J}$ become IGC states if the driving Hamiltonian
assumes the form $\hat{H}_D=\frac{\gamma}{2}(n_+-n_-)\sqrt{s c}(s-c)\hat{S}_y$
with $\hat{S}_y=\sum_{n=1}^N \sigma_{ny}$. 

States with the same $n_+$ have the same eigenvalue and form an IGC
subspace. The dimension of the IGC subspace is given by the binomial coefficient:
$\begin{pmatrix}N\\n_+\end{pmatrix}
$. For $N\geq3$,
there exist IGC subspaces with dimension greater than 1 and can be used to
encode quantum information. For example, for $n=3$ the IGC subspace
corresponding to the eigenvalue $\sqrt{s c}$ can be spanned from the non-orthogonal basis
states: 
\begin{align}
|\psi_1\rangle=&-\sqrt{c}(|101\rangle+|011\rangle)+\sqrt{s}(|100\rangle+|010\rangle)\nonumber \\
|\psi_2\rangle=&-c\sqrt{c}|111\rangle+s\sqrt{c}|100\rangle)-c\sqrt{s}|011\rangle+s\sqrt{s}|000\rangle)\nonumber \\
|\psi_3\rangle=&\sqrt{c}(|011\rangle-|110\rangle)-\sqrt{s}(|001\rangle-|100\rangle).
\end{align} 
The encoding efficiency of the IGC subspace (the number of logical qubits, $\log_2$(Dim(IGC($N$)))) approaches unity just as in the case of the usual DFS \cite{Zan97prl, Lid98}. 

These examples seem to suggest that in finite-dimensional systems IGC states
do only appear in relatively exotic situations. However, it is
quite simple to find realistic examples in  infinite systems.
We give two examples below, noting that since Theorem 6 does not apply to infinite dimensional
systems, these states may not be invariant under time evolution with $H$, despite the preservation
of the state purity.

\begin{example}  Driven damped harmonic oscillator\label{ex:ddho}\end{example}
Consider a driven harmonic oscillator with annihilation operator 
$\hat{a}$ and
\begin{eqnarray} 
  L_D[\rho] &=&  \frac{\gamma}{2}\left[
  2\hat{a}\rho\hat{a}^\dagger-\hat{a}^\dagger\hat{a}\rho-\rho
  \hat{a}^\dagger\hat{a} \right]
\\
  \hat{H}_\text{eff} &=& \omega_0(t) \, \hat{a}^\dagger \hat{a} +
  g^*(t) \, \hat{a} + g(t) \, \hat{a}^\dagger \; .
\label{eq:dho}
\end{eqnarray} 
This model is frequently used to describe photons in a lossy
single-mode cavity that is coherently pumped by a driving field of amplitude $g(t)$.
A coherent state $|\alpha\rangle$ is an eigenstate of 
$\hat{a}$ with eigenvalue $\alpha$ and fulfills
$L_D[|\alpha\rangle\langle\alpha|]\neq 0$ for $\alpha\neq 0$. 
In absence of a driving field ($g(t)=0$) and for constant frequency $\omega_0$, 
this state remains a coherent state \cite{Breuer}.
This also holds in the general case; for a given function $g(t)$ and initial
condition $\alpha(0)$, a coherent state evolves as the pure state
$\rho(t) = |\alpha(t)\rangle\langle\alpha(t)|$ with 
\begin{equation} 
  \alpha(t)= \alpha(0) e^{F(t)} -i \int_0^t \text{d}t'\;
  e^{F(t)-F(t')} g(t')\; ,
\end{equation} 
and $F(t) \equiv \int_0^t \text{d}t' (-i\omega_0(t') -\gamma/2 )$.
Hence, purity is preserved and a coherent state corresponds to an
IGC state. 

As we noted before, Theorem \ref{th:dfs2} does not apply to
infinite-dimensional systems. Thus, purity is preserved for coherent states,
although they are not invariant under the evolution Hamiltonian.  

For a coherent state, the effect of the decoherence term is reduced to an
attenuation of the oscillator: it describes loss of energy 
(or photons) but not loss of coherence.
However, if the choice of $g(t)$ is consistent with the transformation introduced in Eq.~(\ref{eq:dec}), 
for instance in the case 
$\omega_0(t)=\omega_0$ and $g(t)=\frac{i}{2}\gamma\alpha \text{e}^{-i\omega_0 t}$,
then purity and mean photon number $\langle\hat{a}^\dagger\hat{a}  \rho(t) \rangle$
are both preserved, i.e., such a state does not experience decoherence or attenuation.

\begin{example} Two-photon absorber \end{example}

A two-photon absorber pumped by a two-photon parametric process 
constitutes another example of IGC states. The master equation for this system is 
\begin{equation}
\dot{\rho}=-i[\hat{H}_D, \rho]+\frac{\gamma}{2}(2\hat{a}^2\rho\hat{a}^{\dagger
  2}-\hat{a}^{\dagger 2}\hat{a}^2\rho-\rho\hat{a}^{\dagger 2}\hat{a}^2) 
\end{equation}
with $\hat{H}_D=\frac{i\gamma}{2}((\alpha^\ast)^2\hat{a}^2-\alpha^2\hat{a}^{\dagger 2})$.
The two coherent states $|\pm \alpha\rangle$ are solutions to this
master equation \cite{Gerry93, Hach, Gilles94, Guerra97} and are both
eigenstates of the two-photon annihilation operator $\hat{a}^2$ with the same 
eigenvalue $\alpha^2$. Hence they form a two-dimensional IGC subspace. 

\section{Summary and Discussion}
We have examined two different definitions of states $\rho(t)$ of a quantum 
system that undergo unitary evolution in the presence of decoherence effects,
both of which are related to the preservation of state purity.
For the restricted DFS
introduced in definition \ref{def:DFS-1}, the decoherence
term in the master equation (\ref{eq:master}) vanishes, $L_D[\rho(t)]=0$.
This is a sufficient but not a necessary criterion for unitary evolution.
General DFS states can be characterized by the less stringent
condition, 
$ \partial_t\text{Tr}[\rho^2(t)]=0 \text{ for all }t\geq 0$,
or equivalently, $\langle L_\text{D}[\rho(t)]\rangle=0$, that was introduced in
definition \ref{def:gDFS}. With theorems \ref{th:dfs} and \ref{th:dfs2}
we have established rigorous conditions for the existence of 
dynamically stable DFS of both types in finite-dimensional systems.  
We also provided a simple constructive protocol for explicit computation of 
states in all such DFS that exist.

DFS states that fulfill $\langle L_\text{D}[\rho(t)]\rangle=0$
but for which $L_D[\rho(t)]\neq 0$ are especially interesting. These states satisfy Definition~\ref{def:gDFS} but not~\ref{def:DFS-1}. We have
shown that these states rely on incoherent generation of coherences (IGC). This
means that
they correspond to a set of states on which the decoherence term
$L_D[\rho(t)]$ acts like a Hamiltonian term, generating coherences with
other states that are cancelled out by the system Hamiltonian.  We showed that existence of these IGC states constitutes a general phenomenon that can be understood in
terms of basic properties of the Markovian master equation.  In particular, their existence was
shown to result from an invariance of the master equation with respect to a continuous transformation,
implying a dynamical symmetry that links the coherent and decoherent terms and that results in cancellation of the decoherent terms in certain circumstances.

 IGC states form a subset of DFS states
and as such may be useful for quantum information processing.  A general strategy to generate such states is to first identify states on which the decoherence superoperator acts coherently for infinitesimal times and then to stabilize these states to make IGC states by adding a Hamiltonian to the master equation that cancels the action of $L_D[\rho(t)]$ on the states.  We have studied under which decoherence conditions a system admits IGC states.  
For finite-dimensional systems
the most common decoherence models, including independent dephasing and incoherent 
excitation or de-excitation processes, do not allow for IGC states. In other words,
the set of restricted DFS is then identical with the full DFS. IGC states 
in finite-dimensional systems can only appear in relatively exotic 
situations, for instance when
the system interacts with a reservoir that is prepared in a squeezed
vaccum state. On the other hand, in infinite-dimensional systems
it is easy to find examples of IGC states, e.g., coherent
states in a lossy single-mode optical cavity.  

\acknowledgements 
We thank D.~Lidar, A.~Shabani, and P.~Zanardi for valuable critical comments on a
previous version of this work and D. Lidar for helpful comments on this manuscript.  We also thank H. Wiseman for a valuable discussion.
Financial support by iCORE,
 NSERC, and CIFAR is gratefully acknowledged. 
R.I.K. and K.B.W. thank the NSF for financial support under ITR Grant 
No. EIA-0205641, and the Defense Advanced Research Projects
Agency (DARPA) and the Air Force Laboratory, Air Force
Material Command, USAF, under Contract No. F30602-01-
2-0524.


\appendix
\section{Equivalence between unitary dynamics and purity preservation}
\label{app:pure} 
Here we show that Definition \ref{def:gDFS}, according to which a 
state is DF if its purity does not change 
from initial unit value, is equivalent to
requiring unitary dynamics for Markovian master equations. 
We start with the fact that the evolution of any quantum state $\rho_0$ is
always described by a completely positve trace-preserving linear map
(CP map) ${\cal E}_t(\rho_0)$ . For such a map the following
proposition holds.
\begin{Proposition}
\label{prop:unitary}
Let $\rho_0$ be a density matrix with purity one, i.e.,
$ \text{Tr}(\rho_0^2) =1$. Then the action of a linear CP map
${\cal E}_t(\rho_0)$ is unitary, 
\begin{equation} 
  {\cal E}_t(\rho_0) = U_t \rho_0 U^\dagger_t \; ,
\end{equation} 
iff ${\cal E}_t(\rho_0)$ preserves purity, Tr(${\cal E}_t(\rho_0)^2$)=1.
\end{Proposition}

\begin{proof}
We first show that the condition is sufficient. Assume that
${\cal E}_t(\rho_0) = U_t \rho_0 U_t^\dagger$. Then
\begin{eqnarray} 
  \text{Tr}({\cal E}_t(\rho_0)^2) &=& \text{Tr}(U_t\rho_0 U_t^\dagger U_t \rho_0 U_t^\dagger )
\\
  &=&  \text{Tr}(U_t\rho_0 \rho_0 U_t^\dagger )
\\
  &=&  \text{Tr}(\rho_0 \rho_0 )
\\ &=& 1 \; .
\end{eqnarray} 

To prove that the condition is necessary, suppose that 
$\text{Tr}({\cal E}_t(\rho_0)^2)=1$ for a given pure state $\rho_0$.
${\cal E}_t(\rho_0)$ can be written in the operator sum representation as \cite{Kraus}
\begin{equation} 
  {\cal E}_t(\rho_0) = \sum_k \hat{E}_k \rho_0 \hat{E}_k^\dagger \quad , \quad \sum_k \hat{E}_k^\dagger \hat{E}_k = \hat{1}\; .
\end{equation} 
For a state of purity one, the CP map on the pure state $\rho_0 = |\psi \rangle \langle \psi|$  can then be written as
\begin{equation} 
  {\cal E}_t(|\psi \rangle \langle \psi|) = 
  \sum_k |\psi_k \rangle \langle \psi_k|  \quad , \quad
  |\psi_k \rangle \equiv \hat{E}_k |\psi \rangle \; .
\label{eq:CPmap2}
\end{equation} 
This is a uniform mixture of states $|\psi_k \rangle$. Such a mixture can only be of purity one if
all $|\psi_k \rangle$ are proportional to each other, i.e., if
\begin{equation} 
  |\psi_k\rangle = \lambda_k |\tilde{\psi}\rangle \; ,
\label{eq:psik}\end{equation} 
where $|\tilde{\psi} \rangle $ is a normalized state defined as 
$|\tilde{\psi} \rangle = E_1 |\psi \rangle / || E_1 |\psi \rangle ||$, provided that
$|| E_1 |\psi \rangle || \neq 0$. If this is not the case, the normalized state can be defined in terms of any other
$\hat{E}_k$ for which $|| \hat{E}_k |\psi \rangle || \neq 0$. There must be at
least one such operator $\hat{E}_k$, since if there is no such operator, then ${\cal E}_t(\rho_0)
=0$, which would contradict the assumption that purity is preserved.
Using Eq.~(\ref{eq:psik}) in Eq.~(\ref{eq:CPmap2}), we then obtain 
\begin{eqnarray} 
  {\cal E}_t(|\psi \rangle \langle \psi|) &=& \sum_k \hat{E}_k |\psi \rangle \langle \psi| \hat{E}_k^\dagger
\\
  &=& \sum_k |\lambda_k|^2 |\tilde{\psi} \rangle \langle \tilde{\psi} | 
\label{eq:CPmap3}
\end{eqnarray} 
Now we use the unit normalized state $|\tilde{\psi} \rangle $ to write
\begin{eqnarray} 
  \sum_k |\lambda_k|^2  &=&  \sum_k  \langle \tilde{\psi} |  \lambda_k^*  
  \lambda_k |\tilde{\psi} \rangle 
\\ &=&  \sum_k  \langle \psi | \hat{E}_k^\dagger   \hat{E}_k |\psi\rangle 
\\ &=&  \langle \psi | \hat{1} |\psi\rangle 
\\ &=& 1\; ,
\end{eqnarray} 
so that Eq.~(\ref{eq:CPmap3}) becomes
\begin{eqnarray} 
  {\cal E}_t(|\psi \rangle \langle \psi|) &=&  |\tilde{\psi} \rangle \langle \tilde{\psi} |. 
\end{eqnarray} 
Hence ${\cal E}_t$ maps $|\psi \rangle $ to $|\tilde{\psi} \rangle $. Such a map between unit normalized pure states can
always be written as a unitary transformation $U_t$ and explicit construction of $U_t$
can be made, e.g., by using a Householder reflection \cite{lehoucq96:_comput}.  Hence
\begin{equation} 
  {\cal E}_t(|\psi \rangle \langle \psi|) = U_t |\psi \rangle \langle \psi| U_t^\dagger \; .
\end{equation} 
\end{proof}

Under the usual assumption of and initially factorizable system and environment state, the time evolution of a system state $\rho (t)$ can always be described
by a time dependent CP map ${\cal E}_t (\rho_0)$ acting on the initial
state $\rho_0$ \cite{Kraus}.  By proposition \ref{prop:unitary}, requiring unitary dynamics
for all $t$ is then equivalent to the condition 
$\text{Tr} (\rho^2(t))=1\,\forall\, t\geq 0$. For dynamics that are
governed by a Markovian master equation (\ref{eq:master}), $\rho (t)$
is a differentiable operator-valued function. Hence the time derivative of $\rho(t)$ is
well-defined and purity preservation can be expressed as
$\partial_t\text{Tr} (\rho^2(t))=0\,\forall\, t\geq 0$ with
$\text{Tr} (\rho_0^2)=1$. Therefore,  for Markovian dynamics,
Definition \ref{def:gDFS} is equivalent to requiring a DF state to
undergo unitary dynamics.

\end{document}